\documentstyle[psfig]{l-aa}
\def\eta{et al.}
\def\ergs{${\rm erg\,cm^{-2}\,s^{-1}}$ }

\def\mlum{${\rm erg\,s^{-1}\,Hz^{-1}}$ }
\def\mlume{${\rm erg\,s^{-1}\,Hz^{-1}}$}
\def\lo{${\it l}_{\rm o}$ }
\def\loe{${\it l}_{\rm o}$}
\def\lx{${\it l}_{\rm x}$ }
\def\lxe{${\it l}_{\rm x}$}

\def\mlo{$\overline{\it l}_{\rm o}$ }
\def\mloe{$\overline{\it l}_{\rm o}$}
\def\mlx{$\overline{\it l}_{\rm x}$ }
\def\mlxe{$\overline{\it l}_{\rm x}$}

\def\sigmo{$\sigma_{\rm o}$ }
\def\sigmoe{$\sigma_{\rm o}$}
\def\sigmx{$\sigma_{\rm x}$ }
\def\sigmxe{$\sigma_{\rm x}$}
\def\alpox{$\alpha_{\rm ox}$ }
\def\alpoxe{$\alpha_{\rm ox}$}
\def\malpox{$\overline{\alpha}_{\rm ox}$ }
\def\malpoxe{$\overline{\alpha}_{\rm ox}$}
\def\sigmox{$\sigma_{\mbox{\alpoxe}}$ }
\def\sigmoxe{$\sigma_{\mbox{\alpoxe}}$}

\def\ze{{\it z}}
\def\R{$R_{\sigma}$ }
\def\Re{$R_{\sigma}$}
\def\mOLF{$\overline\phi_{\rm o}(\overline{l}_{\rm o},z)$ }

\def\mXLFe{$\overline\phi_{\rm x}(\overline{l}_{\rm x},z)$}
\begin{document}
\thesaurus{03.13.6 -- 11.17.3 -- 13.25.3}
\title{Does the optical--to--X-ray energy distribution of quasars depend on 
optical luminosity?}
\author{W. Yuan \and J. Siebert \and W. Brinkmann}
\offprints{W. Yuan}
\institute{Max--Planck--Institut f\"ur extraterrestrische Physik,
Giessenbachstrasse, D-85740 Garching, Germany}
\date{Received October 8, 1997; accepted January 15, 1998}
\maketitle
\markboth{W. Yuan et al.: Optical--to--X-ray luminosity relationship of quasars}{}
\begin{abstract}
We report on a detailed analysis of the  correlation
between the optical-UV and X-ray luminosities of quasars
by means of Monte Carlo simulations,
using a realistic luminosity function.
We find, for a quasar population with an intrinsically
constant,  mean X-ray loudness \malpoxe,
that the simulated \alpoxe\,--\,$L_{\rm o}$ relation can exhibit
various `apparent' properties,
including an increasing \malpox with $L_{\rm o}$,
similar to what has been found from observations.
The determining factor for this behavior
turns out to be the relative strength of the dispersions of
the luminosities, i.e.\ their
deviations from the mean spectral energy distribution at
the  optical and X-ray bands,
such that a dispersion larger for the optical luminosity than for
the X-ray luminosity tends to result in an apparent correlation.
We suggest that the observed  \alpoxe\,--\,$L_{\rm o}$ correlation
can be attributed, at least to some extent,
to such an effect, and is thus  not an underlying physical property.
The consequences of taking into account the luminosity dispersions 
in an analysis of the observed luminosity correlations 
is briefly discussed.   
We note that similar considerations might also
apply for the Baldwin effect. 
\end{abstract}

\section{Introduction}

A study of the dependence of the spectral energy distribution
 (SED) of quasars  on their luminosity
 and/or on cosmic epoch is particularly important for
understanding the quasar phenomenon.
 In the optical-to-X-ray regime, the SED can
be characterized by the broad band spectral index between 2500{\AA}
and 2\,keV, which is defined as
 $\alpha_{\rm ox} = - 0.384 \log (L_{\rm2keV}/L_{\rm2500{\AA}})$.
 Quasars are known to exhibit strong luminosity evolution
in the X-ray and the optical wave bands (e.g.\ Boyle 1994).
 However, there have been controversial discussions in the past as to whether the
evolution law is the same in these two energy bands.
A dependence of \alpox on
redshift or optical luminosity would indicate different evolution in the
optical and the X-ray regime.
Further, if \alpox depends on optical luminosity, this
is equivalent to a non-linear relationship between X-ray and optical
luminosity ($L_{\rm x}\propto L_{\rm o}^{\rm e}$, $e \not=1$).

While most of the analyses agree on the result that \alpox is redshift independent, 
it has been claimed  that \alpox increases with $L_{\rm o}$, which
implies that the objects with high optical luminosities are under-luminous
in X-rays compared to their low luminosity counterparts (Avni \& Tananbaum 1982,
1986; Kriss \& Canizares 1985; Wilkes \eta\ 1994; Avni \eta\ 1995; Green \eta\ 1995). 
Generally, for a functional dependence of the form 
\alpoxe\,$\sim \beta \log L_{\rm o}$, a canonical slope of $\beta\sim 0.1$ was obtained, 
which is equivalent to a non-linear relation of the form 
$L_{\rm x} \propto L_{\rm o}^{0.7}$.

Based on Monte Carlo simulations, Chanan (1983) suggested that a non-linear
relation might arise even for an intrinsically  linear dependence from 
observational flux limits and the large intrinsic scatter in the data.
Thus, the observed \alpoxe\,--\,$L_{\rm o}$
correlation should not be considered as an underlying physical reality.
The author also claimed that the choice of $L_{\rm o}$
as the independent variable is not justified.
However, Kriss \& Canizares (1985) criticized these results by pointing
out that they depend critically on the assumption
of a Gaussian distribution for the luminosity functions.

A study by La~Franca \eta\ (1995) reinforced the idea of a linear relationship
between the X-ray and the optical luminosity for quasars.
They applied a regression
algorithm to a large sample of quasars detected with {\it Einstein},
which accounts for errors in both variables and
the intrinsic scatter
in the data, and found $L_{\rm x} \propto L_{\rm o}$.

In a recent study of ROSAT detected quasars by Brinkmann \eta\ (1997),
it has been shown by means of a simple argument that an apparent correlation
between \alpox and $\log L_{\rm o}$ can indeed emerge even for
intrinsically uncorrelated variables. Motivated by this idea, as well as by
recent improvements concerning the shape of the quasar luminosity functions
in the optical and the X-ray regime, we carry out a detailed study of this
controversial problem by means of a Monte Carlo analysis. We mostly use the
logarithms of luminosities  and denote them as 
$\mbox{\lxe}\,=\log L_{\rm x}$ and $\mbox{\loe}\,=\log L_{\rm o}$.
We use $q_{\rm 0} = 0.5$, $H_{\rm0} = 50~{\rm kms^{-1}Mpc^{-1}}$
throughout this paper. All errors quoted are at the $1 \sigma$ level unless
mentioned otherwise.

\section{Analysis of luminosity correlations}

\subsection{Distribution of quasar luminosities}

We first describe the assumed parametric form of the multivariate distribution
$\psi$(\loe, \lxe, \ze), which we use in the simulations.
In previous studies
(e.g.\ Avni \& Tananbaum 1982, 1986; Kriss \& Canizares 1985)
$\psi$(\loe, \lxe, \ze)
is commonly expressed as the product of the luminosity function (LF hereafter)
of the primary luminosity (either \lx or \loe) and the conditional
distribution function of the secondary luminosity. The conditional
distribution function determines the expected value of the secondary
luminosity by assuming a functional dependence between the two luminosities
and intrinsic dispersion. In previous models, the dispersion of \alpox was
attributed to the dispersion in the secondary luminosity alone.
In this paper we use a generalized description of the multivariate
distribution function. We assume that both luminosities show intrinsic
dispersion instead of the secondary luminosity only. 

We introduce {\em intrinsic} optical and X-ray luminosities \mlo and \mlxe. 
The observed luminosities \lo and \lx are the intrinsic luminosities
modified by various mechanisms which produce a large scatter.  	
The luminosities \mlo and \mlx are distributed according to their respective
luminosity functions, \mOLF and \mXLFe. 
Although \mlo and \mlx are not directly observable,
their ratio, i.e.\ the {\em intrinsic} \malpoxe, which is defined as
\begin{equation}
\mbox{\malpoxe} \equiv
-\frac{\mbox{\mlo}-\mbox{\mlx}}{\log (\nu_{\rm2500\AA}/\nu_{\rm2keV})}
= 0.384(\mbox{\mloe}-\mbox{\mlxe}),
\label{eq:A}
\end{equation}
can be approximated by the mean of the observed \alpox distribution.   
As indicated by the observed correlation between \lx and \loe,
\mlx and \mlo are physically related.
Assuming a redshift-independent relationship of the form
\begin{equation}
\mbox{\mlx}=f(\mbox{\mloe}) \propto e\,\mbox{\mlo},
\label{eq:B}
\end{equation}
we get an {\em intrinsic} dependence of \malpox on \mloe, 
\begin{equation}
\mbox{\malpoxe(\mloe)}
=\beta_{\rm int} \mbox{\mlo} + \mbox{const},
\quad \mbox{where } \beta_{\rm int}=0.384(1-e). 
\label{eq:C}
\end{equation}
If $e=1$, we have $\beta_{\rm int}=0$, i.e.\
\malpox is independent of the luminosities {\em intrinsically}. 

The {\em observed} optical and X-ray luminosities of a quasar are then
\begin{equation}
l_{\rm o} = \overline{l}_{\rm o} + \delta\l_{\rm o},\qquad
l_{\rm x} = \overline{l}_{\rm x} + \delta\l_{\rm x},
\label{eq:E}
\end{equation}
where $\delta\l_{\rm o}$ and $\delta\l_{\rm x}$ quantify the scatter 
around the intrinsic luminosities in the optical and the X-ray band, respectively. 
The dispersion of \alpox can now be attributed to the dispersions in both \lx and
\loe. Assuming that $\delta\l_{\rm o}$ and $\delta\l_{\rm x}$ are independent,
we have\footnote{
In the general case
$(\delta\mbox{\alpoxe})^2=0.384^2\,[e^2(\delta\l_{\rm o})^2+(\delta\l_{\rm x})^2]$
with $e$ from Eq.\,\ref{eq:B} and $e=1-\beta_{\rm int}/0.384$.
If $e \sim 1$ then $\beta_{\rm int}\sim 0$. The same applies for \sigmox 
in Eq.\,\ref{eq:H}.
}
\begin{equation}
(\delta\mbox{\alpoxe})^2=0.384^2\,[(\delta\l_{\rm o})^2+(\delta\l_{\rm x})^2],
\label{eq:F}
\end{equation}
where $\delta\mbox{\alpox}$ is the dispersion of \alpoxe.

Based on this generalized scenario, the multivariate distribution function
$\psi(l_{\rm o}, l_{\rm x}, z)$ can be replaced by a distribution function 
depending on 
\ze, \mloe, \mlxe, $\delta\l_{\rm o}$ and $\delta\l_{\rm x}$, which is given by
\begin{eqnarray}
\lefteqn{\psi^*(\mbox{\mloe,\mlxe}, \delta\l_{\rm o},\delta\l_{\rm x},z) =}
\nonumber\\
 & & \overline\phi_{\rm o}(\overline{l}_{\rm o}, z)\,
\delta(\overline{l}_{\rm x}-f(\mbox{\mloe}))\,
g_{\rm o}(\delta\l_{\rm o}\,|\,\overline{l}_{\rm o}, z)\,
g_{\rm x}(\delta\l_{\rm x}\,|\,\overline{l}_{\rm o}, z),
\label{eq:G}
\end{eqnarray}
where $\overline\phi_{\rm o}$ is the luminosity function for \mloe, 
$\delta(\overline{l}_{\rm x}-f(\mbox{\mloe}))$ is a 
$\delta$-function with $f$ from Eq.\,\ref{eq:B}, and $g_{\rm o}$  and
$g_{\rm x}$ are the conditional distribution functions for $\delta\l_{\rm o}$
and $\delta\l_{\rm x}$, respectively. In the following we describe
these terms as they were implemented in the Monte Carlo analysis.

We assume that both luminosity dispersions, $\delta\l_{\rm o}$ and
$\delta\l_{\rm x}$, are independent of luminosity and redshift,
and that $g_{\rm o}$ and $g_{\rm x}$ are given by Gaussian distributions
with means of zero, and 
standard deviations $\sigma_{\rm o}$ and $\sigma_{\rm x}$, respectively.
Thus, the distribution of \alpox is Gaussian with standard deviation
\begin{equation}
\sigma_{\mbox{\alpoxe}} =
0.384\,({\sigma_{\rm o}}^2 + {\sigma_{\rm x}}^2)^{1/2}.
\label{eq:H}
\end{equation}
$\sigma_{\mbox{\alpoxe}}$ is available from observational data,
ranging from 0.15 to 0.2 (e.g.\ Avni \eta\ 1995, Yuan \eta\ 1998).
We define the parameter \R as the ratio of the standard deviations of the
optical to the X-ray luminosity dispersions,
\[
\mbox{\R} \equiv \frac{\mbox{\sigmo}}{\mbox{\sigmx}}. 
\]
Then, given \R and $\sigma_{\mbox{\alpoxe}}$, 
\sigmo and \sigmx can be determined using Eq.\,\ref{eq:H}.

Another input is the LF 
$\overline\phi_{\rm o}(\overline{l}_{\rm o},z)$,
which is not directly observable.
 Assuming a $z$-dependent power law for
$\overline\phi_{\rm o}(\overline{l}_{\rm o},z)$
and a Gaussian distribution for
$g_{\rm o}$, it can be shown that the LF for the observed luminosity \lo has
the same functional form and evolution
as that for \mlo except close to the low-luminosity cutoff.
 Therefore, it is natural to approximate $\overline\phi_{\rm
o}(\overline{l}_{\rm o},z)$ by the observed optical luminosity function (OLF)
and its evolution.  
If $\delta\l_{\rm o}$ is small, 
$\overline\phi_{\rm o}(\overline{l}_{\rm o}, z)$
reduces to the observed OLF.

We assume pure luminosity evolution for quasars.
Using the functional form of the OLF as given by Boyle (1994),
which was derived from the UVX sample, we parameterize
$\overline\phi_{\rm o}(\overline{l}_{\rm o},z)$
by a broken power law with $\gamma_1 = -1.6$ for
\mloe$< l_{\rm o}^*$ and $\gamma_2 = -3.9$ for \mloe $> l_{\rm o}^*$, and
$\overline{l}_{\rm o}^*(z)= \overline{l}_{\rm o}^*$(\ze=0)$+ k\cdot 
\log(1+z)$, $k\sim 3.5$, for $z<2$.
A low-luminosity cutoff was applied at
$M_{\rm B}^{\rm min} = -20$ at $z=0$ for \mloe.

It should be noted that \mlo and \mlx are equivalent and interchangeable in
this model. Eq.\,\ref{eq:G} can also be expressed in terms of \mlxe, in which case the
X-ray luminosity function (XLF) has to be used instead of OLF.

\subsection{Monte Carlo analysis}

The Monte Carlo analysis was performed by generating a sample of quasars
with {\it z}, \loe, and \lx as follows:\\
First, a redshift {\it z} was drawn from a given range $(z_1, z_2)$
satisfying $V(z_1,z)/V(z_1,z_2)=r_1$, where
$V(z_1,z)$ and $V(z_1,z_2)$ are the volumes within $(z_1, z)$
and $(z_1, z_2)$, respectively,
and $r_1$ is a random number between 0 and 1.
The volume element in co-moving space is given by (for $q_0=0.5$)
\begin{equation}
dV(z)= 16\pi (c/H_0)^{3} (1+z)^{-3.5} (1+z-\sqrt{1+z})^2\,dz.  
\label{eq:I}
\end{equation}
Then, a second random number $r_2$ was calculated
and the intrinsic luminosity at the optical band
$\overline{l}_{\rm o}$ was determined
such that
$\int^{\overline{L}_{\rm o}}_{\overline{L}_{\rm o}^{\rm min}}
\overline\phi_{\rm o}(\overline{L^{'}}_{\rm o},z)\,d\overline{L^{'}}_{\rm o}\,/\,
\int^{\tiny \infty}_{\overline{L}_{\rm o}^{\rm min}}
\overline\phi_{\rm o}(\overline{L^{'}}_{\rm o},z)\,d\overline{L^{'}}_{\rm o}
=r_2$, with the luminosity function
$\overline\phi_{\rm o}(\overline{L}_{\rm o},z)$
as specified above.
After that, the expected X-ray luminosity $\overline{l}_{\rm x}$ 
was obtained using Eq.\,\ref{eq:A} for a given \malpoxe.
The {\it observed} optical and X-ray luminosities, \lo and \lxe, are then
drawn from the Gaussian distribution functions 
    $g_{\rm o}(l_{\rm o}-\mbox{\mloe})$
and $g_{\rm x}(l_{\rm x}-\mbox{\mlxe})$, respectively. 
The corresponding fluxes
were calculated from the luminosities assuming $\alpha_{\rm o} = -0.5$ and
$\alpha_{\rm x} = -1.3$ for the K-corrections, respectively. An object was
accepted, if the fluxes were above the given flux limits. In general we
simulated optically selected samples assuming a threshold magnitude 
$m_{\rm B,th}$ = 20.
If the objects turned out to have an X-ray flux below the
assumed X-ray flux limit they were treated as `non-detections'.

The above procedure was repeated until a sample comprising 300 objects
was  obtained.
For each object, \alpox was calculated from the simulated \lo and
\lxe. We then investigated the \alpoxe\,--\,\lo relationship by means of a
Spearman rank correlation test and least-square linear regression analysis.
For a specific set of parameters of the quasar population, we carried out 10
independent trials and acquired 10 samples. All statistical quantities
derived below are the mean of the individual values of the 10 samples,
along with the statistical uncertainty for the mean.

\subsection{Results}

As the first and simplest case
we assumed the intrinsic \malpox and \sigmox in Eqs.\,\ref{eq:C} and \ref{eq:H} 
to be constant, i.e. independent of luminosity. In accordance with 
observational results we used
\malpoxe\,$= 1.40$  and  \sigmoxe\,$= 0.18$.
We considered 20 values of the \Re-parameter from \Re$= 0.1$ to $10$,
which are sampled evenly on a logarithmic scale.
Eq.\,\ref{eq:H} then determines the dispersions \sigmo and \sigmxe.    
We investigated optically selected samples with a size of $N = 300$
and a limiting magnitude $m_{B \rm ,th} = 20$.
In a first attempt, we restrict our analysis to samples
at a fixed redshift ($z=1$), for which the results give an insight
into the physical problem,
and then we present cases for samples in larger redshift ranges.

\subsubsection{Sample at fixed redshift}
The Spearman correlation coefficients $\rho_{\rm sp}$
are plotted in Fig.\,1 as open squares for 20 different \R values,
with the typical $1 \sigma$ error indicated for illustration.
The probability levels $P_r = 0.01$ and 0.001,
at which the `no correlation' hypothesis is ruled out,
are shown at the corresponding $\rho_{\rm sp}$ ($N =300$) as dotted
and dashed lines, respectively.
\begin{figure}
\psfig{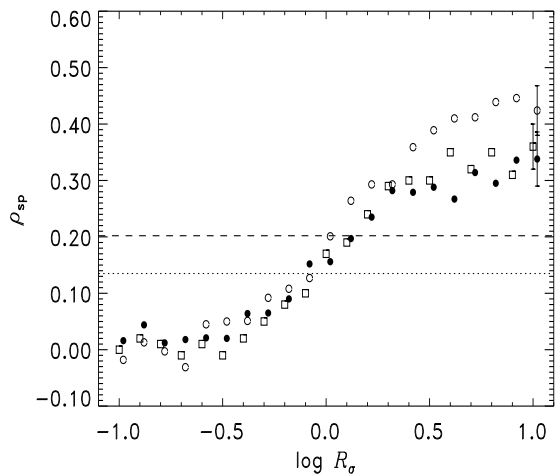}
\caption[]{Spearman rank correlation coefficients $\rho_{\rm sp}$ as a measure
of the \alpoxe\,--\,\lo correlation for simulated samples for various 
\Re-parameters; squares: sample at fixed redshift;
filled circles: sample with redshift $0.2 < z < 3$;
open circles: sample with redshift $0.2 < z < 3$ and containing only X-ray 
detections. The dotted and dashed lines indicate the $\rho_{\rm sp}$
for which the `no correlation' hypothesis can be ruled out
at the corresponding probability levels (one tail)
of $0.01$ and $0.001$, respectively.}
\end{figure}
Despite of the luminosity independence of the intrinsic mean \malpox
the results show the emergence of a positive correlation
with increasing \Re,
which becomes significant for \Re\,$>1$.
In Fig.\,2 we show an example of  an \alpox versus \lo plot 
for a simulated sample with \Re\,$= 5$.
The dashed line represents a constant X-ray luminosity  (\lxe\,=\,27.2\,\mlume).
Obviously, the simulation indicates a much sharper cutoff in X-ray luminosity
compared to optical luminosity, which has already been noted by Brinkmann 
\eta\ (1997) for observed data.

\begin{figure}
\psfig{figure=0702.f2,height=7.0cm,width=8.0cm,angle=0}
\caption[]{\alpox versus \lo for a simulated quasar
sample at $z = 1.0$, assuming a constant \malpox and \Re\,$= 5$.
The lines indicate  constant X-ray luminosities,  
\lxe\,$= 27.2$\,\mlum (dashed) and \lxe\,$= 26.5$\,\mlum (dotted).}
\end{figure}

Now we consider the effects of thresholds in the X-ray observations.
At a redshift of $z$=1 a 
luminosity threshold of $l_{\rm x}^{\rm th} \sim 26.5$\,\mlum at 2\,keV
indicated by the dotted line in Fig.\,2 
corresponds to a flux limit of
$\sim 4\times10^{-14}$\,\ergs in the 0.5--2\,keV band.
Objects below this line have $l_{\rm x} \geq l_{\rm x}^{\rm th}$ and
can thus be `detected' in X-rays;
whereas the rest with  \lxe\,$< l_{\rm x}^{\rm th}$ are `non-detections'.
When only the `detections' are considered\footnote{Hereafter, 
we refer to such a case as `with an X-ray threshold',
otherwise we refer to data `without an X-ray threshold'},
the X-ray thresholds can enhance the apparent
\alpoxe\,--\,\lo correlation significantly, as can be inferred from the figure. 

The results are independent of redshift,
which changes only the range of optical luminosities.

\subsubsection{Samples in redshift ranges}
We now consider samples with a redshift range of $0.2 < z < 3.0$.
The results of the correlation analysis are added to Fig.\,1 for the
case without (filled circles) and with an X-ray threshold (open circles).
We find almost the same results as for the sample at fixed redshift.

The slopes $\beta$  obtained by fitting 
a linear \alpox $\sim \beta\cdot\l_{\rm o}$
relation to the simulated samples, are plotted in Fig.\,3 for various \R values,
for the case without (filled circles) and with (open circles) X-ray thresholds, 
respectively.
The results are fully consistent with those of the rank correlation test.
Another effect to be noted is that the resulting average
\alpox of the obtained sample
is consistent with the intrinsic \malpox ($= 1.4$)
only for small \R ($\ll 1$),
and it increases towards higher \R with a typical difference
of $\sim 0.1$ for \Re$\gg 1$.
\begin{figure}
\psfig{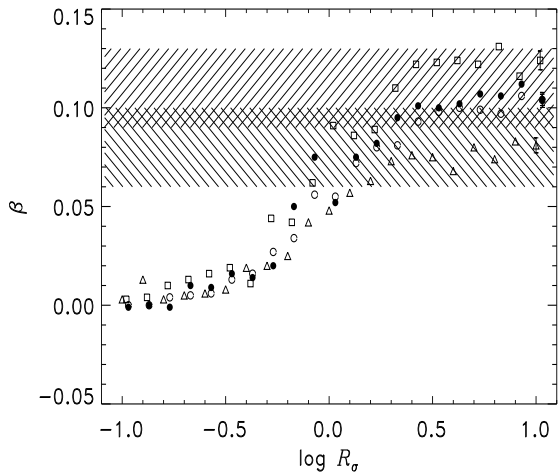}
\caption[]{Fitted slopes of \alpoxe\,$\sim \beta \log L_{\rm o}$
for various \R values. Filled circles: no X-ray threshold considered and
\sigmoxe\,$=0.18$; open circles: with X-ray threshold and \sigmoxe\,$ = 0.18$;
triangles: with X-ray threshold and \sigmoxe\,$=0.15$; squares: with X-ray 
threshold and \sigmoxe\,$=0.20$. The two shaded areas indicate the typical 
range of slopes ($1\,\sigma$) given in two  previous studies (see text).}
\end{figure}

The fitted slopes are found to be independent of the specified
value of \malpox  and insensitive to the
X-ray observational thresholds as well as to sample completeness.
However, they depend slightly on the dispersion \sigmox
(within plausible limits on \sigmoxe) in a manner that,
at large \Re, the slope increases with \sigmox for a given \Re.
We plot in Fig.\,3 the results for samples
assuming \sigmoxe\,$=0.15$ (triangles) and \sigmoxe\,$=0.2$ (squares), respectively,
ignoring the effect of X-ray thresholds.
The shaded areas in the figure indicate the $1\sigma$ confidence region
of $\beta$ obtained in previous studies
($\beta_{\rm obs} = 0.11 \pm 0.02$, Wilkes \eta\ 1994, Yuan \eta\ 1998;
$\beta_{\rm obs} = 0.08 \pm 0.02$, Green \eta\ 1995).
It shows that the observed slopes can be reproduced
within a relatively large region in the parameter space of \R and \sigmoxe.
Furthermore, the slopes also depend only marginally  on
the low-luminosity cutoff $M_{\rm B}^{\rm min}$ chosen for \mloe,
which is somewhat uncertain.

In Fig.\,4 we show the simulated \alpox and \lo values 
for \Re\,$= 5$. For `non-detections' lower limits on \alpox are shown as crosses.
This distribution, as well as that for samples at fixed redshift (Fig.\,2),
differs from what is expected for a linear \alpoxe\,$\sim$\,\lo relation, 
showing asymmetric structures with respect to a regression line.
\begin{figure}
\psfig{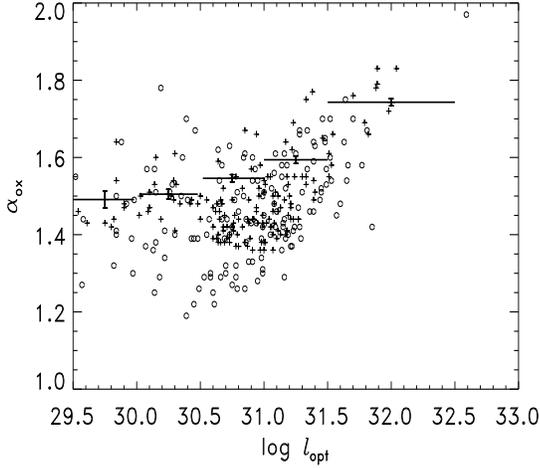}
\caption[]{\alpox versus $\log L_{\rm o}$ for a simulated quasar sample
in the redshift range 
$0.2 < z < 3 $, assuming \Re\,=\,5. Crosses indicate lower limits for objects
which would not have been detected in the presence of an observational
X-ray  threshold.
Thick bars indicate the best estimates of the mean \alpox in \lo bins
which were obtained by taking into account non-detections (see Sect.~3.2).}
\end{figure}

To test the correctness of the simulation procedure 
each sample was subjected to a self-consistency check using
the evolution weighted Schmidt's (1968) $V/V_{\rm m}$, as in Chanan (1983).
The simulated samples represented the envisaged luminosity function 
evolution as the resulting $\langle V/V_{\rm m} \rangle$
were distributed randomly around 0.5 within the $1~\sigma$ uncertainties.

We also performed similar analyses starting from \mlx instead of \mloe.
After having determined $z$, \mlx was drawn from 
the LF for \mlxe,
$\overline\phi_{\rm x}(\overline{l}_{\rm x},z)$,
which is approximated by the observed XLF.
We used the broken power law LF as given in Boyle (1994),
with a faint-end slope $\gamma_1 = -1.6$,
a bright-end slope $\gamma_2 = -3.3$,
and evolution rate $k=3.3$ for $z<2$.
Then, a corresponding \mlo was obtained from Eq.\,\ref{eq:A}
for the given \malpoxe.
The observed luminosities were drawn in the same way as above.
Similar results were found, i.e.\ an apparent \alpoxe\,--\,\lo
correlation towards high \Re.
This is expected because of the striking
similarity between the OLF and XLF (Boyle 1994).

We conclude that an apparent \alpoxe\,--\,\lo correlation can
emerge even for a population with an intrinsically constant \malpox
in the presence of large dispersions in the SED and for \Re\,$\geq 1$.
The degree of correlation depends on the \Re-parameter.
Similar results hold for X-ray selected quasar samples, as well as for 
incomplete samples.

\section{Discussion}

\subsection{Intuitive consideration}
The simulation results can be understood in terms of simple intuitive 
arguments. We show schematic sketches for the \lxe\,--\,\lo relation in Fig.\,5a 
and the corresponding \alpoxe\,--\,\lo relation in Fig.\,5b for a sample at 
fixed redshift. Following Sect.~2.1, the distribution of luminosities in the 
optical and X-ray regime is determined by the mean \malpoxe.
A constant \malpoxe\,$= 1.4$ is assumed (thick lines in Fig.\,5a and b),
as well as constant luminosity dispersions \sigmo and \sigmx (indicated for 
an object at Q). The expected luminosities \mlo and \mlx are distributed 
along the line of constant \malpox according to their distribution function
$\overline\phi_{\rm o}(\overline{l}_{\rm o}(\overline{l}_{\rm x}), z)$,
with a cutoff at the bright end of \mlo and \mlx for a sample with finite size.

\begin{figure}
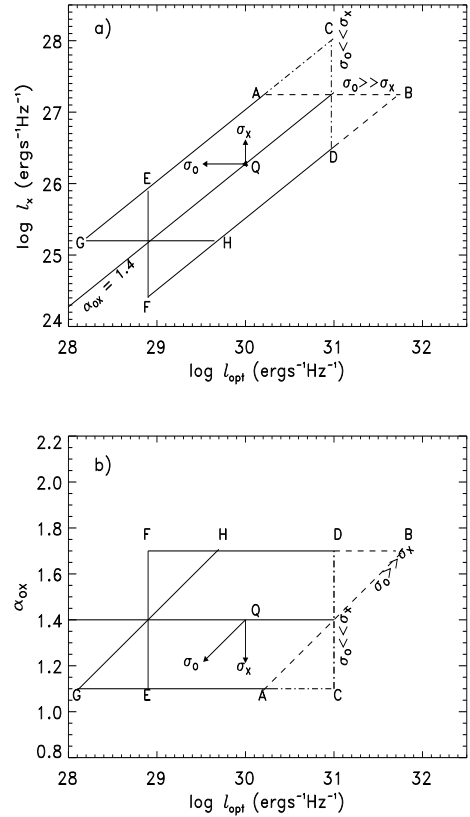

\psfig{figure=0702.f5a,height=5.7cm,width=7.0cm,angle=0}
\psfig{figure=0702.f5b,height=5.7cm,width=7.0cm,angle=0}
\caption[]{Schematic sketches for the \lxe\,--\,\lo and \alpoxe\,--\,\lo 
relationship. See text for a detailed explanation of the various symbols.}
\end{figure}

We consider the distribution of objects
deviating from \mlo and \mlx due to luminosity scatter
in the \lxe\,--\,\lo and \alpoxe\,--\,\lo planes.
The 90\% probability region for an object is confined
within the two thin lines parallel to the mean \malpoxe.
We consider two extreme cases: \sigmoe\,$\gg$\,\sigmx (\Re\,$\gg 1$)
and \sigmoe\,$\ll$\,\sigmx (\Re\,$\ll 1$).
For \sigmoe\,$\gg$\,\sigmxe,
the luminosity scatter
is predominant in the optical luminosity, i.e.\ along the optical axis.
The distribution of objects in the \lxe\,--\,\lo plane forms a confined region 
below the horizontal dashed line AB, which corresponds to the line AB with 
slope 0.384 in Fig.\,5b. An apparent correlation appears between 
\alpox and \lo for such distributions, despite the intrinsically uncorrelated relation.
This effect explains the simulated \alpoxe\,--\,\lo distribution in Fig.\,2.
On the other hand, in the case of \sigmoe\,$\ll$\,\sigmxe,
objects are distributed in a region confined by line CD (dashed-dotted) and
no \alpoxe\,--\,\lo correlation is expected from the high-\lo end.
The actual degree of correlation increases with \sigmo relative to \sigmxe,
which explains the dependence \alpoxe\,--\,\lo correlation on \Re, as seen
in Figs.\,1 and 3. Moreover, flux limits in X-rays (indicated by the line GH),
tend to enhance the apparent \alpoxe\,--\,\lo correlation, as can be seen from
Fig.\,5b and Fig.\,2. For an optically selected sample, the cutoff due to the optical flux 
limit (indicated by EF) systematically excludes objects with smaller \alpox 
values, which qualitatively explains the resulting increase of average \alpox for large \Re.

We note that, in order to get a sharply confined \alpoxe\,--\,\lo  distribution 
at the high luminosity cutoffs, a steep luminosity function for \mlo (\mlxe)
is required towards the high luminosity end. Our simulations show that the 
observed optical and X-ray quasar luminosity functions satisfy this condition.

Using simulations basically similar to ours and a much steeper luminosity 
function (Gaussian distribution), Chanan (1983) found the existence of a 
\alpoxe\,--\,\lo correlation in his cases B and C, but not in case A, 
corresponding to \Re\,$\sim 1.4$, \Re\,$ = 1$, and \Re\,$ \sim 0.8$ in our analysis.
These values are close to the critical point \Re\,$ = 1$, 
 and the results of Chanan  are thus not 
  a consequence of the reversal of dependent and independent variables
in the regression analysis.
Furthermore, the use of the \Re-parameter in our work, i.e.\
the relative strength of \sigmo and \sigmxe,
is physically more meaningful than the assumption that the 
luminosity of one wave band directly 
determines that in the other band (as in Chanan 1983).
The case of \Re\,$\gg 1$ leads to a  more pronounced high-luminosity
cutoff for \lx compared to \lo (see also Fig.\,2). This effect is also seen
in observational data (Brinkmann \eta\ 1997).

It can also be seen in Fig.\,5a that,
when performing a linear regression analysis for the \lxe\,--\,\lo relation,
different distributions of the data  near the highest luminosities
can result in different slopes for different \Re.
Thus, a simple linear regression analysis method is not always adequate
to quantify a luminosity correlation,
especially for data with large inherent scatter.

\subsection{Dependence of \alpox on \lo --- intrinsic or apparent?}

Our results show that an apparent dependence of \alpox on \lo
can emerge from data for a quasar sample with no intrinsic dependence.
Thus, the explanation of the observed \alpoxe\,$\sim$\,\lo correlation
as a physical relation in quasars,
as taken for granted in  previous work, must  be questioned.
Although the current study does not allow to unambiguously distinguish
between an {\em intrinsic} or an {\em apparent} dependence of \alpox on \loe,
we may reach some conclusions by simple considerations.

If our model is a good description of the quasar luminosity correlation and 
dispersion, we  expect the  existence of an apparent correlation to 
some extent, unless in cases of rather small \R ($\la 0.3$), i.e.\ when the 
dispersion in \lo is at least 3 times less than that in \lxe, which seems 
unlikely considering the diversity of the big-blue-bump component in quasar 
energy spectra.
In fact, the aforementioned characteristics of a much tighter high-luminosity
cutoff for \lx than for \lo seems to exist in most quasar samples of 
considerable size.
Thus, an intrinsic \alpoxe\,--\,\lo dependence, if it does exist,
must be weaker than it  appears from the  data.

To quantify these effects, we carried out similar simulations incorporating
an intrinsic luminosity dependent \malpox as in Eq.\,\ref{eq:C},
$\beta_{\rm int} \neq 0$,
and compare the obtained slopes $\beta$ of the \alpoxe\,--\,\lo relation with the
observed values.
For each grid point in the $\beta_{\rm int} -$\,\R parameter space,
we repeated the simulations for 200 trials
and counted the number $m$  of trials for which $\beta$ fell into
a region around the observed value $\beta_{\rm obs} \pm u_{\rm \xi}\sigma$, at
a confidence level $\xi$.
The overall confidence level of reproducing the observed slope
is thus $\xi \cdot m/200$, if systematic uncertainties, such as sample 
incompleteness etc., are ignored. For $\beta_{\rm obs} = 0.11 \pm 0.02$
(Wilkes \eta\ 1994, Yuan \eta\ 1998), 
Fig.\,6 shows the contours of the 68\% (thick lines) and 
90\% (thin lines) confidence regions of the joint $\beta_{\rm int} -$\,\R 
distribution for \sigmoxe\,=\,0.15 (dotted), 0.18 (solid), and 0.20 (dashed), respectively.
The results show that only a weak dependence ($\beta_{\rm int} \la 0.05$)
is needed for $\mbox{\Re} \ga 1$, when
the dispersion in \lo is comparable to or larger than that in \lxe;
and almost no intrinsic dependence is needed for larger \Re\,$\ga 3$.
It also shows that the most probable (68\%) region
appears at high \R ($\ga 1$), low $\beta_{\rm int} (\la 0.07) $ values. 
\begin{figure}
\psfig{figure=0702.f6,height=7.0cm,width=8.0cm,angle=0}
\caption[]{Contours of the 68\% (thick lines) and 90\% (thin lines)
confidence region in \R and $\beta_{\rm int}$, the slope of the intrinsic 
\alpoxe\,--\,\lo dependence, for \sigmoxe\,=\,0.15 (dotted), \sigmoxe\,=\,0.18 (solid)
and \sigmoxe\,=\,0.20 (dashed).}
\end{figure}

More qualitative evidence comes from some features of the 
observed \alpoxe\,--\,\lo relation.
Recent studies of large quasar samples
(Avni \eta\ 1995, Brinkmann \eta\ 1997, Yuan \eta\ 1998)
have shown that the observed \alpoxe\,--\,\lo correlation shows a more complex behavior,
with only weak or even vanishing correlation at low optical luminosities.
We find that such features are  expected,  
as the apparent correlation for \Re\,$>1$  arises mainly from objects with high 
optical luminosities. 
This property has been verified by estimating
the average \alpox in different \lo bins for the simulated data.
This can be seen 
in Fig.\,4, where we plot the average \alpox (thick bars) for five \lo bins,
which were obtained by incorporating upper limits for the
`non detections' using the maximum likelihood method developed by Avni \eta\ (1980).

A comparison of the observed OLF and XLF and their evolution
may also give, in principle,
constraints on the intrinsic \malpoxe\,--\,\mloe, or \mlxe\,--\,\mlo relation
(Eqs.\,\ref{eq:B} and \ref{eq:C}).
For a power law luminosity function  for $\overline{L}_{\rm o}$
with index $\gamma_{\rm o}$,
and luminosity evolution  $ \sim (1+z)^{k_{\rm o}}$,
it can be shown (see also Kriss \& Canizares 1985)
that the corresponding LF for $\overline{L}_{\rm x}$
is also a power law (index $\gamma_{\rm x}$) with evolution $(1+z)^{k_{\rm x}}$ and
\begin{equation}
\gamma_{\rm x}-1 = (\gamma_{\rm o}-1)/e, \qquad 
k_{\rm x} = e k_{\rm o}.
\end{equation}
For a Gaussian approximation of the luminosity dispersion,
the luminosity functions 
$\overline\phi_{\rm o}(\overline{l}_{\rm o}, z)$ , 
$\overline\phi_{\rm x}(\overline{l}_{\rm x}, z)$
have the same forms and evolution as the observed luminosity functions 
$\phi_{\rm o}(l_{\rm o},z)$, $\phi_{\rm x}(l_{\rm x},z)$. 
Thus, $e$ and $\beta_{\rm int}$ can be estimated from the 
observed $k_{\rm o}$, $k_{\rm x}$, $\gamma_{\rm o}$ and $\gamma_{\rm x}$.

The optical evolution index is found to be $k_{\rm o}= 3.45 \pm 0.1$ (Boyle 1994). 
However, the X-ray evolution index 
is somewhat uncertain---ranging from $k_{\rm x}=2.56\pm0.17$
(Della Ceca \eta\ 1992, $q_0=0$, $Einstein$ EMSS data only),
$k_{\rm x}=3.0^{+0.2}_{-0.3}$ (Jones \eta\ 1997) and
$k_{\rm x}=3.34\pm0.1$ (Boyle \eta\ 1994) to
$k_{\rm x}=3.2 - 3.5$ (Franceschini \eta\ 1994).
The latter three values were obtained by incorporating ROSAT data,
and they are consistent with $e = 1$ and $\beta_{\rm int} = 0$ within their 
2 $\sigma$ errors, i.e.\ no intrinsic dependence of \alpox on \lo is needed.
Even if we take the smallest value of $k_{\rm x}=3.0$ of Jones \eta\ (1997),
we have $e=0.87$ and $\beta_{\rm int}=0.05$,
implying only a marginal dependence,
weaker than the previous claims of $\beta \sim 0.1$.
The bright-end slopes of the XLF ($\gamma_2 = 3.30^{+0.34}_{-0.09}$)
and of the OLF ($\gamma_2 = 3.9\pm 0.1$, Boyle 1994) are
consistent with each other within their 2$\sigma$ errors,
and thus also consistent with $e = 1$ and $\beta_{\rm int} = 0$.
We conclude that a constant \malpox  independent of \lo is not
ruled out by a comparison of  the observed luminosity functions in the 
optical and the X-ray band.

\section{Conclusions}

We have performed Monte Carlo simulations to study
the luminosity correlation between the optical and X-ray bands for quasars.
We have used a generalized model 
in which the luminosities in the two wave bands are
represented in terms of the respective expected luminosities with dispersions
($l_{\rm o} = \overline{l}_{\rm o} + \delta\l_{\rm o}$,
$ l_{\rm x} = \overline{l}_{\rm x} + \delta\l_{\rm x}$).
We have shown that the increase of \alpox with $L_{\rm o}$ (equivalent to
$L_{\rm x} \propto L_{\rm o}^e$ with $e<1$),
as found in observational data, can emerge in a sample with an intrinsic 
luminosity independent \alpox  (or $L_{\rm x} \propto L_{\rm o}$),
provided that the dispersion of the optical luminosities
deviating from the average SED are similar to or larger than that of
the X-ray luminosities. Our simulations  verified the results of  Chanan 
(1983), which were achieved for special assumption about the luminosity functions.
We suggest that the {\em observed} \alpoxe\,--\,$L_{\rm o}$ correlation is,
at least to a large extent, apparent 
and not necessarily an intrinsic property of the quasar population.

Our model is more general than  previous considerations.
For $\mbox{\Re}\ll 1$ (implying $l_{\rm o} \sim \overline{l}_{\rm o}$ and
$\sigma_{\mbox{\alpoxe}} \sim 0.384 \sigma_{\rm x}$)
the model reduces to the commonly used description, in which
\lo is the primary luminosity and the dispersion in the SED is
attributed to that in the X-ray luminosity. The same  holds for 
$\mbox{\Re}\gg 1$, but with interchanged roles of \lo and \lxe.
We argue that the effect of the relative strength of the individual luminosity 
dispersions in the two bands should be taken into account in 
analyses of quasar luminosity correlations. 
Since the arguments are valid for any other two wave bands,
we expect this effect to play a role in luminosity correlations between
radio, infrared, optical, and X-ray wave bands as well.
The determination of the \Re-parameter, and thus of
the luminosity scatter in the individual wave bands,
is important to understand
the broad band emission of quasars.

We finally note that a similar effect as presented
for the $\alpha_{\rm ox} - l_{\rm o}$ correlation might also
apply for the well-known Baldwin effect, i.e.\ the inverse
correlation of optical emission line equivalent width with optical 
luminosity (Baldwin 1977). Since the equivalent width basically 
can be regarded as the ratio of two luminosities (emission line and 
underlying continuum), the structure of the problem is similar to 
the one presented in this paper. This is particularly interesting,
because there still is no accepted physical explanation for the 
Baldwin effect.


\end{document}